\documentstyle[preprint,prl,aps,epsf]{revtex}

\begin{document}

\draft

\title{Crossover from percolation to diffusion }

\author{D. N. Tsigankov, A. L. Efros}

\address{\vspace{-0.1cm}Department of Physics,  University of Utah, Salt Lake City UT 84112 USA}

\maketitle



\begin{abstract}
A  problem of the crossover from percolation to diffusion transport is considered. A general scaling theory is proposed. It introduces phenomenologically four critical exponents 
which are connected by two equations.
 One exponent is completely new.
 It describes the increase of the diffusion
below percolation threshold.
As an example, an exact solution
 of one dimensional lattice problem is given.
 In this case the  new exponent $q=2$.

\end{abstract}

Percolation theory is often used to describe transport properties of
 disordered systems with
 a large disorder. The typical problems are random  mixture of conducting
 and non-conducting elements, or hopping conduction\cite{rev}.
 The regular percolation
 theory assumes that the random
 elements do not change their positions
 with time so that the percolation paths
 do
 not change their spatial
 locations. The low-temperature electron transport is one of examples where
 this
 assumption is not valid. Due to electron-electron interaction the random potential persistently
 and  substantially changes with time\cite{ef,men}, which may affect conductivity near
 the metal-insulator transition.  Such problems appear
 outside solid state physics as well. The class of this problem is known as
 {\em dynamical percolation}. They have been studied theoretically
 using effective medium  approximation (See e.g. Ref.\cite{avik} and references
 therein). One dimensional problems of this type have been considered for
 some models  without any approximations\cite{step}.

We concentrate here on the  case when the  diffusion
 through conducting media is drastically faster  then the  fluctuations
 of the positions of the conducting and non-conducting elements.
 In this case
the transition from diffusion to percolation mechanism has all features of
 the phase transition and it can be characterized by critical exponents.
 In this paper we introduce a set of these exponents
and establish
 relation between them. For illustration we present an exact
 solution of one dimensional lattice  problem.

Thus, we consider here a random mixture of the conducting and non-conducting
 elements which can change
there positions but very slowly. In this situation the resulting diffusion
 (or conductivity)
 of the particles {\em is non-zero at any small fraction $p$ of
the conducting elements}.
 The mechanism of this diffusion is as follows. A particle can diffuse only
in the conducting medium. To move from one conducting element to the other it
 is waiting
until another conducting element comes to the element where it resides.
 At this moment of time a particle
 is able to do  a next move. This is the simple diffusion which is characterized by some 
slow  waiting time $\tau_s$.
 The diffusion inside conducting medium
 is characterized by  much faster time
$\tau_f$. When the fraction of conducting elements $p$ becomes close to the percolation
 threshold $p_c$, but it is less than $p_c$ the resulting diffusion
increases as
 $1/(p_c-p)^q$,
 where $q$ is a novel
 critical exponent. This happens
 because the conducting  clusters
 become
 large. But since they are disconnected the particle should wait
 to jump  from one
of them to the other.
 Finally, when $p-p_c$ is positive and it becomes
 larger than
 the width of
some critical region, the diffusion is described by a regular percolation law
 $D\sim (p-p_c)^t$
 and slow motion of the conducting elements is not important.
To describe this physics one can use the same scaling arguments as for
 the problem of frozen
mixture of elements with large and small conductivity\cite{efr,str}. Note
 that the new
 problem is not equivalent to the  old one, so that the exponent $q$ might
 be different.
The scaling hypothesis is valid in  the proximity of the percolation
 threshold so that $|X|/p_c\ll 1$, where
$X=p-p_c$. In this region the scaling  hypothesis can be written in the form
\begin{equation}
d=h^s\psi({X\over h^m}).
\label{scale}
\end{equation}
Here  $d=D/D_f$, $h=D_s/D_f$, where $D_f=a^2/2\tau_f$ is the diffusion coefficient in 
the conducting medium,$a$ is a characteristic length which depends   on the model, 
$D_w=a^2/2\tau _s$, and $\psi(Z)$ is some analytical function at
 all real values of its argument, $-\infty <Z< \infty$ .
 We assume that $h\ll 1$ and that
$\psi(0)=1$.

  Eq. (\ref{scale}) contains two independent critical exponents. The meaning of the 
exponent $s$ is  that
$d=h^s$ at
$p=p_c$. The exponent
$m$ describes the width
 of the
critical region $|p-p_c|=h^m$ between the percolation and diffusion.

 All the critical
 exponents which describe $d(X)$ can be related to $s$ and $m$.
At $X>0$ and $X\gg h^m$ the slow changes of the percolation paths
 are not important. Thus, we have the diffusion on
 the percolation cluster. If the system is larger than the correlation radius
of the percolation cluster,
 the diffusion coefficient is
related to the conductivity by the Einstein relation. Then one has
$d\sim X^t$, where the exponent $t$ is a usual percolation exponent
 which describes the conductivity above the percolation threshold.
To get this result from Eq. (\ref{scale}) one should assume  that
$m=s/t$.
 At $X<0$ and
$|X|\gg h^m$ we expect that
$d\sim h/(|X|)^q$. This gives $s+mq=1$. Finally we get two  connections between four exponents
 $s,q,t,m$, namely $q=t(s^{-1}-1)$ and $m=s/t$.

 As an example,  we consider below an exactly soluble one dimensional site
 or bond  model
on a lattice. At the beginning we consider the site model. The sites may be
 white and black  with the fractions $p$
  and $1-p$ respectively.  The particle may occupy the white sites only and it is able to  jump at
   the nearest site, if this site is white, during the time $\tau_f$.
   The configuration of the
  white and black sites slowly changes  with time. This change can be introduced by many 
  ways.
  We consider the simplest one. After each time interval $\tau_s$, which is
   called the renewal time,
  the configuration of all sites completely changes preserving the same $p$, while the
   particle remains at the same white site. We assume  that
   $\tau_s\gg  \tau_f$.
  Since  $p_c=1$ in the one-dimensional case, our solution may
   only illustrate the increase of the
   diffusion coefficient $D\sim (1-p)^q$
   and the width of the critical region $|X|$, where it deviates from this law
  and tends to the value $D_f=a^2/2\tau_f$ which is reached at  $p=1$.
  Since the diffusion of  a particle within the time interval $\tau_s$ is completely
   independent of
   the diffusion during other time intervals, one has

  \begin{equation}
  \overline{r^2(t)}=\overline{(r_1(t_1)+r_2(t_2)+.
  ..+r_N(t_N))^2}=\overline{r_1^2(t_1)}+\overline{r_2^2(t_2)}+...+\overline{r_N^2(t_N)}
  \end{equation}

  It follows that

  \begin{equation}
   D= \lim_{t\rightarrow\infty}\frac{r^2(t)}{2t}
  =\frac{\overline{r^2(\tau_s)}}{2\tau_s},
  \end{equation}

  where
  $\overline{r^2(\tau_s)}$ is  averaged over all possible initial position of
   a particle.

  Now our strategy is as follows.
   We assume first that the clusters of the white
   sites are not very large so that the particle, which change position during
  the time $\tau_f$ crosses them to and fro  many
  times during the time $\tau_s$. In this approximation we find
   diffusion coefficient $ D_1$. For the larger clusters we use the continuum approximation, 
  which  is valid at $1-p\ll 1$ only, and find $ D_2$. This coefficient gives the
  correct value at $p=1$. These two approximations   have a large region of overlap if
  \begin{equation}
   {\tau_f\over \tau_s} \ll 1.
  \label{in}
  \end{equation}
   Matching $D_1$ and $D_2$ in the overlap region  we find the result which
   is exact  if Eq. (\ref{in}) is fulfilled.

  To calculate $D_1$ we assume that the particle crosses each cluster many
   times  during the time $\tau_s$ .Then the average quadratic displacement
   $R^2(S)$ is independent of $\tau_s$ and
   it is given by the equation

  \begin{equation}
  R^2(S) 
  =\frac{a^2}{S^2}\sum_{n=1}^{S}\sum_{k=1}^{S}(n-k)^2=\frac{a^2}{6}(S^2-1).
  \end{equation}

  The probability that a particle is within a
   cluster of $S$ white sites is
  \begin{equation}
  w_s=(1-p)^2p^{S-1}S.
  \label{dis}
  \end{equation}

  By averaging  $R^2(S)$  over all the
   clusters one gets

  \begin{equation}
  \overline{r^2(\tau_s)}=\frac{a^2}{6}((1-p)^2\sum_{S=1}^{\infty}p^{S-1}S
  (S^2-1)=\frac{a^2p}{(1-p)^2}.
  \end{equation}
  Thus,
  \begin{equation}
  D_1=\frac{\overline{r^2(\tau_s)}}{2\tau_s}=\frac{D_sp}{(1-p)^2},
  \label{d1}
  \end{equation}
  where
  $D_s=a^2/2\tau_s$.
  Note that  Eq. (\ref{d1}) has been obtained
  by Druger {\it et al.}\cite{step}.
  The Eq.(\ref{d1}) is valid if
   $\overline{r^2(\tau_s)}<<D_f \tau_s$. It is fulfilled if
  \begin{equation}
   \frac{p}{(1-p)^2}\ll \frac{\tau_s}{\tau_f}.
  \label{con1}
  \end{equation}

  At $p\ll 1$ one gets
  \begin{equation}
  D_1={a^2p\over \tau_s}.
  \end{equation}
  To calculate $D_2$ we use the continuous approximation which is valid when the size of
   white  clusters $S\gg 1$.
  This is true if $1-p\ll 1$. In this approximation one  should solve
   the diffusion equation

  \begin{equation}
  \frac{\partial{u(x,t)}}{\partial{t}}= D_f\frac{\partial^2{u(x,t)}}{\partial{x^2}}
  \end{equation}

  assuming zero current at the beginning and at the
  end  of the cluster ($x=0,L$)
  \begin{equation}
   \frac{\partial{u(x,t)}}{\partial{x}}|_{x=0,L} = 0.
  \end{equation}
   The initial condition has a form
  \begin{equation}
  u(x,t)|_{t=0} = \delta (x-x_0),
  \end{equation}
  where $x_0$ is a random point within the interval $(0,L)$.

  The solution has a form
  \begin{equation}
  u(x,t)=\frac{1}{L}+\sum_{n=1}^{\infty}\cos\left(\frac{\pi n 
  x_0}{L}\right)\cos\left(\frac{\pi n x}{L}\right)\exp\left(-\frac{\pi^2 n^2}{L^2}{ 
  D_f}t\right).
  \end{equation}
   The mean squared displacement with respect  to the initial position $x_0$ is

  \begin{eqnarray}
  &\displaystyle \overline{r_L^2(x_0,t)}=\int_0^L(x-x_0)^2 u(x,t) dx=
   \nonumber \\
  &\displaystyle 
  =\frac{L^2}{3}-(L-x_0)x_0+4L^2\sum_{n=1}^{\infty}\frac{\cos\left(\frac{\pi n 
  x_0}{L}\right)}{(\pi 
  n)^2}\left[\frac{x_0}{L}+(-1)^n\left(1-\frac{x_0}{L}\right)\right]\exp\left(-\frac{\pi^2 
  n^2}{L^2}D_ft\right)
  \label{disp1}
  \end{eqnarray}

  Finally, taking the average over the initial positions $x_0$ on the cluster, we get  the time 
  dependence of
   the mean square
   displacement of the particle on the cluster of the size $L$.

  \begin{eqnarray}
  &\displaystyle \overline{r_L^2(t)}=\frac{1}{L}\int_0^L\overline{r^2(x_0,t)} 
  dx_0=\nonumber \\
  &\displaystyle 
  =\frac{L^2}{6}-L^2\sum_{n=0}^{\infty}\frac{1}{\left(\frac{\pi}{2}(2n+1)\right)^4}\exp\left(-\frac{\pi^2 
  (2n+1)^2}{L^2}D_ft\right)=
  \nonumber \\
  &\displaystyle 
  =\frac{16}{\pi^4}L^2\sum_{n=0}^{\infty}\frac{1}{(2n+1)^4}\left[1-\exp\left(-\frac{\pi^2 
  n^2}{L^2}D_ft\right)\right].
  \label{disp}
  \end{eqnarray}

  One can see from Eq. (\ref{disp}) that for relatively small  times,  $t << L^2/D_f$, the 
  displacement
    $\overline{r_L^2(t)}$ grows as $2D_ft$, just as for the normal diffusion,
  while at large times the value  $\overline{r_L^2(t)}$  tends
   to the asymptotic value  $L^2/6$ indicating that the particle crosses
  the cluster many times.

  To find the diffusion coefficient $D_2$ one has   to average $\overline{r_L^2(\tau_s)}$  
  over  all clusters
  with the  distribution function given by Eq. (\ref{dis}).

   In
   the region $(1-p)<<1$ one  can substitute summation over the cluster sizes by 
  integration to get
  \begin{eqnarray}
  &\displaystyle 
  \overline{r^2(\tau_s)}=\frac{(1-p)^2}{a^2}\int_0^{\infty}\overline{r_L^2(\tau_s)}L\exp(\frac{L}{a}\ln{p}) 
  dL=\nonumber \\
  &\displaystyle 
  =\frac{a^2}{(1-p)^2}-\frac{16}{\pi^4a^2}(1-p)^2\sum_{n=0}^{\infty}\frac{1}{(2n+1)^4}\int_0^{\infty}L^3\exp\left(\frac{L}{a}\ln{p}-\frac{\pi^2 
  n^2 a^2}{2L^2}\frac{\tau_s}{\tau_f}\right)dL
  \end{eqnarray}

  The diffusion coefficient $D_2$ can be represented in a form

  \begin{eqnarray}
  &\displaystyle{D_2}= \frac{\overline{r^2(\tau_s)}}{2\tau_s}= \frac{16 
  D_s}{\pi^4a^4}(1-p)^2\sum_{n=0}^{\infty}\frac{1}{(2n+1)^4}\int_0^{\infty}L^3\exp\left(\frac{L}{a}\ln{p}\right)\left[1-\exp\left(-\frac{\pi^2 
  n^2 a^2}{2L^2}\frac{\tau_s}{\tau_f}\right)\right]dL
  \end{eqnarray}
  or
  \begin{eqnarray}
   D_2=\frac{16D_s}{\pi^4(1-p)^2}\sum_{n=0}^{\infty}\frac{1}{(2n+1)^4}\int_0^
  {\in
  fty}x^3\exp(-x)\left[1-\exp\left(-\frac{\pi^2(2n+1)^2}{2x^2}\frac{\tau_s}{\tau_f}(\ln{p})^2\right)\right]dx
  \end{eqnarray}
  where $x=-\frac{L}{a}\ln{p}$.

  Introducing  the dimensionless parameter $A=\pi^2(\ln{p})^2 \tau_s /2\tau_f$ and
   changing the integration variable $x=b_n\sqrt{z}$, where $b_n=\sqrt{A}(2n+1)$ one 
  can perform the summation over
  $n$ to get
  \begin{equation}
  D_2=\frac {a^2A}{\pi^2 \tau_f}\int_0^{\infty} {z(1-\exp(-1/z))dz\over
   \sinh(\sqrt{A z})}.
  \label{int1}
  \end{equation}
  At $A\gg 1$ one can neglect the exponent in the nominator of Eq.(\ref{int1}). Then

  \begin{equation}
  D_2=D_s/(1-p)^2
  \label{lar}
  \end{equation}
  At $A\ll 1$ one can expand this exponent to get $D_2= D_f$. One can also
  get the next term at small  $A$. Thus,
  \begin{equation}
  D_2=D_f\left[1-\frac{2\sqrt{A}}{\
  pi^2}\int_0^{\infty}\sqrt{z}\left(\frac{1}{z}-1+\exp\left(-\frac{1}{z}\right)\right)dz\right].
  \label{int}
  \end{equation}
  Calculating the integral in Eq. (\ref{int}
  ) one gets
  \begin{equation}
  D_2= D_f(1-0.479\sqrt{A})= D_f(1-1.06\sqrt{\tau_s/\tau_f}(1-p))
  \end{equation}
  at $A\ll 1$.

  Note that the linear dependence in $1-p$ means that the correction is proportional
  to the ratio of the mean displacement  of a particle $\sqrt{D_f \tau_s}$
   during the time $\tau_s$ to the typical size of a white cluster $(1-p)^{-1}$.

   The  scaling arguments in  some different form than Eq. (\ref{scale}) can be applied to 
  the one-dimensional case
   as well. It can be written in a form
  \begin{equation}
  D_2=D_f F(\frac{1-p}{h^m}),
  \end{equation}
  where $h=\tau_f/\tau_s$. It has the same form as Eq. (\ref{scale}) except that
   $s=0$. This relation is defined at $p\le 1$ and $1-p\ll 1$. In the same way as before,
   we put  $F(0)=1$. At large values of $(1-p)/h^m$ one has $D\sim h/(1-p)^q$.
   It follows that $mq=1$ which is analogous to the relation $s+mq=1$ in
   the many-dimensional case. It follows from Eq.(\ref{lar})
   that $q=2$.  At $1-p \ll 1$ one can write $A=\pi^2(1-p)^2 \tau_s /2\tau_f$.
  Then it follows that $m=1/2$
   in
  agreement with scaling relation. The exponent $t$ is not defined in
   the one-dimensional case.

  Finally we present the solution for the effective diffusion coefficient $D$
   which is valid at all values of $p$ within the interval $0\le p\le 1$. It has been shown that 
  $D=D_1$ if  $p/(1-p)^2\ll\tau_s/\tau_f$. On the other hand
  $D=D_2$ if $(1-p)\ll 1$. Thus, the two approximation have a wide region of
  overlap $(p\tau_f/\tau_s)^{0.5}\ll (1-p)\ll 1$. In this region
   $D_1=D_sp/(1-p)^2$ and $D_2=D_s/(1-p)^2$. Thus one can get a result which is
   exact everywhere if $\tau_s\gg\tau_f$. This result is $pD_2$. Finally,

  \begin{equation}
  D=\frac {a^2Ap}{\pi^2 \tau_f}\int_0^{\infty} {z(1-\exp(-1/z))dz\over
   \sinh(\sqrt{A z})}.
  \label{res}
  \end{equation}
  This result has been derived above for the site problem. One can see, however,
  that it remains unchanged for the bond problem as well.

  The diffusion coefficient $D/D_f$ as given by  Eq.(\ref{res}) is plotted in Fig. 1 versus 
  $(1-p)$ in the double
  logarithmic scale at different  values of
  $\tau_s/\tau_f$. The function $(1-p)^{-2}$ is also shown there as a dot-dashed
     line. One can see that at large values of
  $\tau_s/\tau_f$ the  curves have  wide regions which  are  parallel to this line. In these 
  regions the
  diffusion coefficient increases as
  $(1-p)^{-2}$.

  Finally, we have presented a novel problem which fills a gap between
   diffusion and percolation in case when the motion of the random media is very
  slow. We have
  considered the dc transport only. It would be interesting to  study a frequency dependent 
  transport in the same
  conditions.

  We thank B. D. Laikhtman for a fruitful discussion. This work was supported by the 
  US-Israel Binational Science
  Foundation, grant 9800097.

  \begin{figure}[htbp]

  \centerline{\epsfxsize=7.5in\epsfbox{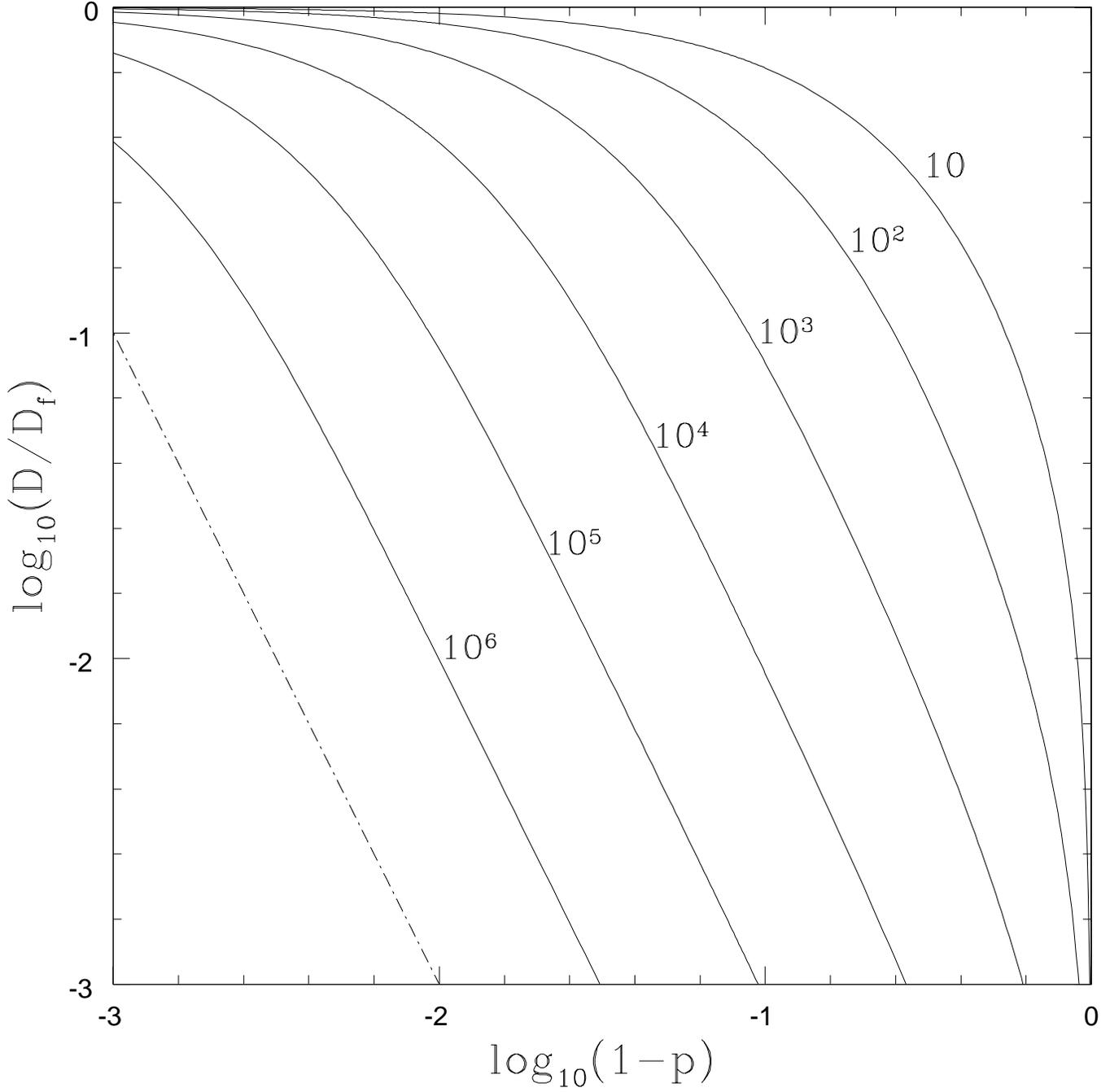}}

  \caption{The  function $D/D_f$ is plotted versus $1-p$ at different values of 
  $\tau_s/\tau_f$ in the double logarithmic scale. The function
  $ (1-p)^{-2}$ is shown by dot-dashed staight line. }

  \end{figure}

  \end{document}